\documentclass{caosp309}

\usepackage{graphicx}

\usepackage{natbib}
\articleNo{NNN}
\pubyear{2024}
\volume{NN}
\volnumber{NN}
\firstpage{NN}
\received{MM, DD, YYYY}
\accepted{MM, DD, YYYY}

\def\BibTeX{{\rm B\kern-.05em{\sc i\kern-.025em b}\kern-.08em T\kern-.1667em\lower.7ex\hbox{E}\kern-.125emX}}

\begin{document}

%
\hauthor{J.\,Southworth and P.F.L.\,Maxted}

\title{The PLATO Multiple Star Working Group (MSWG)}


%
%
\author{J.\,Southworth\inst{1}\orcid{0000-0002-3807-3198} \and P.F.L.\,Maxted\inst{1}\orcid{0000-0003-3794-1317}}

%
\institute{Astrophysics Group, Keele University, Staffordshire, ST5 5BG, UK \\ \email{taylorsouthworth@gmail.com}}

\date{Month Day, Year}

\maketitle

\begin{abstract}
The PLATO mission is scheduled for launch in December 2026. It is an ESA M-class mission designed to find small planets around bright stars via the transit technique. The light curves it obtains will be wonderful for other science goals, among which is the study of binary and multiple stars. We are creating the Multiple Star Working Group (MSWG) to bring together the community to best exploit this unique opportunity. We will assemble the many science cases, create target lists, and co-ordinate applications for PLATO observations. We include instructions on how to register your interest.
\end{abstract}

\section{The PLATO mission}

PLATO (Planetary Transits and Oscillations of Stars) was selected by the European Space Agency (ESA) in 2014 as a mission to search for small transiting extrasolar planets \citep{Rauer+24xxx}. Its baseline aim is to to detect a 1\,R$_\oplus$ planet orbiting in the habitable zone of a G0\,V star of magnitude $V=10$, and to determine the planetary radius to 3\% and system age to 10\%. The planet detection and characterisation will be performed using the transits, and the age determination will be obtained from asteroseismology.

PLATO will consist of one spacecraft hosting 26 cameras, of two types. The main scientific instruments are 24 ``normal'' cameras (N-CAM), each of which will have a 12\,cm entrance pupil and a 4510$\times$4510-pixel CCD. Each will have a field of view of 1037\,deg$^2$ with a sampling of 15$^{\prime\prime}$ per pixel, and will observe in a red passband (500-1000\,nm). The other cameras are two fast cameras (F-CAM) which will be used for fine-pointing of the satellite and may also be available for high-speed observations. They have smaller fields of view (610\,deg$^2$) and different passbands to the science instruments (505--700\,nm for the blue and 665--1000\,nm for the red camera).

PLATO is scheduled for launch in December 2026, and will take approximately 90 days to move to the second Lagrangian point in the Earth--Sun system. Its observations will be divided into 90-day segments between which the satellite will rotate by 90\degr\ to keep its solar panels illuminated by the Sun. The baseline duration of the mission is four years, of which at least the first two years will be used to observe the LOPS2 (the second Long-duration Observation Phase South) field in the southern hemisphere \citep{Nascimbeni+22aa}. The spacecraft will hold consumables to last 8.5 years and the strategy for the remainder of the mission will be decided after observations of LOPS2 have commenced \citep{Rauer+24xxx}.

\section{Complementary science}

A set of PLATO Complementary Science Work Packages\footnote{\texttt{https://fys.kuleuven.be/ster/research-projects/plato-cs/work\_packages}} have been formed to aid the organisation of all scientific areas which will benefit from the availability of PLATO data but do not contribute to the core science goals. The Complementary Science Work Packages include binary and multiple stars, stellar pulsation and rotation, mass loss, debris discs, galactic structure and transient phenomena, and are co-ordinated by Conny Aerts and Andrew Tkachenko. 

The Complementary Science teams will submit applications for PLATO data using the same Guest Observer (GO) mechanism as other scientists, and will have no influence on the observing strategy for PLATO. The GO programme has been allocated 8\% of the PLATO data rate, and the call for proposals is expected nine months before launch.

\section{WP 161\,000: Binary and Multiple Stars}

Complementary Science Work Package 161\,000 (PI Southworth) deals with binary and multiple stars. Our task is to co-ordinate target selection, GO applications and scientific work within the community. This will be done by forming a PLATO Multiple Star Working Group (MSWG) which all members of the scientific community with suitable research interests can join. Our first goal will be to compile a White Paper detailing the work of the Working Group and thoroughly discussing the science cases which we can address. This will naturally lead into the second goal: to co-ordinate the GO applications.

A catalogue of variable stars will be made available prior to the GO deadline. Work at Keele University will lead to an automated pipeline to estimate the physical properties of eclipsing systems based on data from the NASA Transiting Exoplanet Survey Satellite \citep[TESS; ][]{Ricker+15jatis} and \textit{Gaia} missions (see Overall \& Southworth, these proceedings). The PLATO MSWG will then co-ordinate applications for GO observations. We will be able to select as targets anything within the PLATO Input Catalogue (PIC; \citealt{Montalto+21aa}) and outside the PLATO prime sample. Some targets are already in the process of being selected via the PLATO scvPIC \citep[Science Calibration and Validation PIC; ][]{Zwintz24eas} and WP~125\,500 (PLATO Benchmark stars) activities.

PLATO is expected to produce light curves of extremely high quality, thus continuing the revolution within the field of binary and multiple stars triggered by the previous \textit{CoRoT}, \textit{Kepler} and TESS missions \citep{Me21univ}. The combined differential photometric precision (CDPP) over two hours is less than 100\,ppm for stars brighter than TESS magnitude 12 \citep{Borner+24exa}. This is substantially better than TESS \citep{Eschen+24mn} -- and comparable to \textit{Kepler} \citep{Borucki16rpph} but over a much larger field of view -- in order to satisfy the scientific requirements. These data will be useful for a huge range of science cases including, but not limited to: 
\begin{itemize}
\item determination of the properties of massive stars; 
\item the study of pulsating stars in eclipsing binary systems; 
\item the study of binary systems in stellar clusters and associations;
\item analysis of giant and subgiant stars in eclipsing binaries; 
\item investigation of the radius discrepancy in low-mass stars; 
\item the study of mass exchange and loss in interacting binaries; 
\item assessing magnetic activity in binary stars via detection of starspot and flare activity;
\item the detection of multiplicity through additional sets of eclipses and/or eclipse timing and duration variations;
\item the population of binary and multiple star systems.
\end{itemize}
Many of these activities, in turn, are valuable for furthering the knowledge of the structure and evolution of stars, by constraining phenomena such as convective mixing, tidal effects, angular momentum transport and magnetic activity.

\section{How to get involved}

Anyone interested in joining the PLATO MSWG and helping with this work can contact the PLATO MSWG at the email address \texttt{platomswg@gmail.com} -- please provide a brief statement of your science interests and how you would like to contribute. The MSWG is formally launched with this conference proceedings and will be organised via a webpage, teleconferences, in-person meetings, and the production of a white paper on multiple-star science to aid the forthcoming applications for PLATO GO time. We look forward to hearing from you!


\acknowledgements
\noindent JS and PFLM acknowledge support from the Science and Technology Facilities Council (STFC) under grant number ST/Y002563/1. We are grateful for extensive discussions about PLATO with many colleagues.

\end{document}